\documentclass[]{aastex631}

\begin{document}

\title{A detectable ultra-high-energy cosmic ray outburst from GRB 221009A}

\author[0000-0002-8941-9603]{Hao-Ning He}
\affiliation{Key Laboratory of Dark Matter and Space Astronomy, Purple Mountain Observatory, Chinese Academy of Sciences, Nanjing 210023, China}
\affiliation{Astrophysical Big Bang Laboratory, RIKEN, Wako, Saitama 351-0198, Japan}
\email{hnhe@pmo.ac.cn}

\author[0000-0003-2478-333X]{B. Thoedore Zhang}
\affiliation{Center for Gravitational Physics and Quantum Information, Yukawa Institute for Theoretical Physics, Kyoto University, Kyoto 606-8502, Japan}

\author[0000-0002-8966-6911]{Yi-Zhong Fan}
\affiliation{Key Laboratory of Dark Matter and Space Astronomy, Purple Mountain Observatory, Chinese Academy of Sciences, Nanjing 210023, China}
\affiliation{School of Astronomy and Space Science, University of Science and Technology of China, Hefei 230026, China}
\email{yzfan@pmo.ac.cn}

\begin{abstract}

Gamma-ray bursts (GRBs) have been proposed as one of promising sources of ultra-high-energy cosmic rays (UHECRs), but observational evidence is still lacking. 
The nearby B.O.A.T. (brightest of all time) GRB 221009A, an once-in-1000-year event, is able to accelerate protons to $\sim 10^{3}$ EeV. 
Protons arriving at the Milky Way
are dominated by neutron-decay-induced protons.
The inter-galactic magnetic fields would not yield a sizable delay of the $\geq 10{\rm~EeV}$ cosmic rays if its strength is $\lesssim 10^{-13}{\rm~G}$,
while Galactic magnetic fields would cause a significant time delay. 
We predict that, an UHECR burst from GRB 221009A would be detectable by the Pierre Auger Observatory and the TA$\times$4, within $\sim$ 10 years. 
The detection of such an UHECR outburst will provide the direct evidence for UHECR acceleration in GRBs.

\end{abstract}


\section{Introduction} \label{sec:intro}

The origin of the highest energy cosmic particles is still in debate.
GRBs, the most luminous phenomenon in the universe,
are proposed as one of the most promising sources of UHECRs  \citep{waxman1995,Vietri1995}. 
However, no observed UHECRs or TeV-PeV neutrinos have been confirmed to be associated with GRBs \citep{IceCube2012,IceCube2022}.
One of the difficulties of identifying UHECRs associated with GRBs is the deflections and the consequent time delays caused by the magnetic fields.
The magnetic deflection angles and the delay times are dependent on the distance of GRBs and the properties of the interstellar (both in the host galaxy and the Galaxy) and inter-galactic magnetic fields. 
What's more, UHECRs would lose a good fraction of energy via interacting with the extragalactic background light (EBL) and cosmic microwave background (CMB) photons during the propagation,
implying that distant GRBs can hardly contribute to UHECRs.
In addition, the pointing direction of the GRB jet, the capability of accelerating UHECRs, the total energy deposited into CRs, and the escape rate of UHECRs from the burst and the host galaxy are also key issues for the nearby individual GRB \citep{Meszaros2002,Kumar2015}.

GRB 221009A, locating at $z=0.151$ (the luminosity distance of 745 Mpc\citep{Malesani2023}) and releasing an isotropic-equivalent radiation energy of $E_{\gamma,\rm iso}\simeq10^{55}{\rm ergs}$ \citep{An2023}, is the brightest of all time (B.O.A.T.) \citep{Pillera2022,redshift2022,Castro-Tirado2022,Malesani2023, JWST2023}. Throughout this work, the convention $Q_{\rm x}=Q/10^{x}$ is adopted. 
It is also the first GRB with photons of energy extending up to $\sim$10 TeV \citep{LHAASO2023, LHAASO:2023lkv}, indicating its potential ability of accelerating UHECRs \citep{Mirabal2023,Batista2022,Das2023, Zhang:2022lff, Zhang:2023uei, Isravel:2023thi}.
Likely, only 1 of $10,000$ randomly generated long bursts is as energetic as GRB 221009A, 
and the rate of GRBs as nearby and as energetic as GRB 221009A is $\lesssim 1$ per $1000$ years \citep{Burns2023, Williams2023}.
Therefore, GRB 221009A provides us a precious opportunity to study UHECRs in our life time.

The energy dissipation radius of this GRB is large, i.e., $R=2\Gamma^2ct_{\rm v}/(1+z)=10^{15}~{\rm cm}~(1+z)^{-1}\Gamma_{2.7}^2\frac{t_{\rm v}}{0.082{\rm s}}$, where the variability timescale is measured as $t_{\rm v}=0.082~{\rm s}$ \citep{Liu2023}, based on the {\it Fermi}-GBM data from $210\,{\rm s}$ to $219\,{\rm s}$ after the {\it Fermi}-GBM trigger, and $c$ is the speed of light.
The bulk Lorentz factor $\Gamma$ suggested by LHAASO observations on the afterglow is also large \citep{LHAASO2023}.
The large energy dissipation radius and the large bulk Lorentz factor $\Gamma$ favor a large maximum acceleration energy of protons.
If protons are accelerated at the energy dissipation radius during the prompt emission phase,
the photomeson production interaction and the Bethe-Heitler pair production process between accelerated protons and prompt photons will produce neutrons, neutrinos, and electromagnetic particles initiating electromagnetic cascades (contributing to GeV-TeV photons) \citep{Bottcher1998,Dermer2006,Wang2018}. 
The non-detection of neutrinos by the IceCube observatory, observations on GeV photons by {\it Fermi}-LAT, and observations on VHE photons by LHAASO set constraints on the acceleration and interaction mechanism of protons \citep{IceCube2023,Veres2022,Bissaldi2022,Ai2023,Murase2022,Liu2023,Rudolph2023,Wang2023}.
The tightest constraint is made by \citep{Wang2023} based on LHAASO observations, and the baryon loading factor is constrained to be $\xi_{p}\equiv E_p/E_{\gamma,\rm iso}\leq2$ for a Lorentz factor $\Gamma=500$, suggesting a possible high isotropic energy deposited in protons $(E_p)$.
LHAASO's observations on the afterglow emission suggest that the jet of GRB 221009A is 
pointing to Earth. The fact that the jet pointing to Earth allows the accelerated UHECRs to propagate to the direction of Earth.

In aggregate, the specific features of the B.O.A.T. GRB 221009A, such as the large energy dissipation radius, the large bulk Lorentz factor, the high isotropic energy, and the jet pointing to Earth, favor the burst as a possible UHECR source.  
To answer the question of whether the UHECR burst from GRB 221009A can be detected in our life time, 
we study the acceleration and escape of UHECRs in the burst, and simulate the propagation of UHECRs from the burst. 
Since the jet composition of high-luminosity GRBs is likely to be dominated by protons \citep{BTZ2018}, throughout this work, we assume the composition of UHECRs from the GRB is pure proton.

\section{The acceleration and escape of protons} 
Since most of the radiation energy is released via the prompt emission during the brightest emission phase from $225.024{\rm~s}$ to $233.216{\rm~s}$ after the Konus-WIND trigger time in the second pulse \citep{Frederiks2023}, we assume most of the UHECRs are accelerated during the brightest emission phase in the second pulse as well. Parameters of the prompt emission for the brightest phase in the second pulse are listed in Table \ref{table_parameters}.
Protons are expected to be accelerated at the energy dissipation radius within the dynamic time, and meantime are undergoing energy loss due to the synchrotron emission, the hadronic interaction with prompt photons.

The dynamic timescale is $t'_{\rm dyn}=R/(c\Gamma)\simeq 67{\rm~ s} R_{15}\Gamma_{2.7}^{-1}$.
The magnetic field at the energy dissipation radius can be calculated by assuming the energy fraction of the magnetic field and electrons as $\epsilon_B$ and $\epsilon_{e}$, i.e., $B'=(8\pi \epsilon_B L_{\gamma}/(\epsilon_e4\pi R^2\Gamma^2c))^{1/2}\simeq 5.2\times10^3~{\rm G}~\epsilon_{e,-1}^{-\frac{1}{2}}
    \epsilon_{B,-2}^{\frac{1}{2}}\Gamma_{2.7}^{-1}R_{15}^{-1}L_{\gamma,54}^{\frac{1}{2}}$,
where $L_{\gamma}$ is the calibration luminosity.
The acceleration timescale of protons with energy of $\varepsilon_p$ is $t'_{\rm acc}(\varepsilon_{p})=R'_L(\varepsilon_p)/c\simeq 4.9~{\rm s}~(1+z)\varepsilon_{p,20}\eta_{0}^{-1}\epsilon_{e,-1}^{\frac{1}{2}}
    \epsilon_{B,-2}^{-\frac{1}{2}}R_{15}L_{\gamma,54}^{-\frac{1}{2}}$.
The synchrotron cooling time scale of protons is $ t'_{\rm syn}(\varepsilon_{p})=9m_p^4\Gamma/(4ce^4B'^2\varepsilon_{p}(1+z))\simeq7.3\times10^2~{\rm s}~(1+z)^{-1}\varepsilon_{p,20}^{-1}\epsilon_{e,-1}
    \epsilon_{B,-2}^{-1}\Gamma_{2.7}^{3}R_{15}^{2}L_{\gamma,54}^{-1}$,
where $m_p$ is the rest mass of protons.

The number density of the GRB prompt photons is described by 
$\frac{dn(\varepsilon_\gamma)}{d\varepsilon_\gamma}=A_{\gamma}\left(\frac{\varepsilon_\gamma}{\varepsilon_{\gamma,\rm b}}\right)^{-q}$
where the normalization factor is
$A_{\gamma}\sim  L_{\gamma}/(4\pi cR^2 2\varepsilon_{\gamma,\rm b}^2)$,
with $q=\alpha$ for $\varepsilon_{\gamma}<\varepsilon_{\gamma,\rm b}$, $q=\beta$ for $\varepsilon_{\gamma}>\varepsilon_{\gamma,\rm b}$.
Adopting the $\Delta$ resonance approximation, 
the cross section peaks when the energy of photons is $\varepsilon_{\gamma,\Delta}\simeq 0.3\rm GeV$ in the proton rest frame.
The energy loss timescale of the photomeson production interaction between protons and prompt photons is approximated to be \citep{Waxman1997,He2012}
\begin{eqnarray}\label{eqn:photomeson}
t'^{-1}_{p\gamma}(\varepsilon_{p})&&\simeq
5.3\times10^{-3}{\rm s^{-1}}(1+z)^{-1}L_{\gamma,54}R_{15}^{-2}\Gamma_{2.7}^{-1}\varepsilon_{\gamma,\rm b, 3MeV}^{-1}\\\nonumber
&&\times\left\{
    \begin{array}{ll}
 \left(\frac{\varepsilon_{p}}{\varepsilon_{p,\rm b}}\right)^{\beta-1},\,\,\,\,\varepsilon_p<\varepsilon_{p, \rm b}\\\nonumber   \left(\frac{\varepsilon_{p}}{\varepsilon_{p, \rm b}}\right)^{\alpha-1}.\,\,\,\,\varepsilon_p>\varepsilon_{p, \rm b}
    \end{array} \right. 
\end{eqnarray}
where $\varepsilon_{p, \rm b}\equiv\Gamma^2\varepsilon_{\gamma,\Delta}m_pc^2/(2(1+z)^2\varepsilon_{\gamma,\rm b})\simeq8.9\times10^{15}{\rm eV}\Gamma_{2.7}^{2}(1+z)^{-2}\varepsilon_{\gamma, \rm b,3\rm MeV}^{-1}$ is the characteristic energy of protons that interacting with photons of energy $\varepsilon_{\gamma,\rm b}$ via $\Delta$ resonance.

{ Adopting parameters summarized in Table \ref{table_parameters},} the acceleration timescale, the energy loss timescales, and the dynamic time of protons at the energy dissipation radius are calculated and plotted in Figure \ref{fig:timescales}.
The curve of the energy loss timescale via the photomeson interaction is calculated numerically, since the analytical approximation shown in Eq. (\ref{eqn:photomeson}) is not accurate for protons at the high energy end, where the multi-pion channel plays an important role. 

By comparing the dynamic timescale $t'_{\rm dyn}$, the proton synchrotron cooling timescale $t'_{\rm syn}$, the photomeson production timescale $t'_{ p\gamma}$ with the proton acceleration timescale $t'_{\rm acc}$,
i.e., setting ${\rm min}[t'_{\rm dyn},~t'_{\rm syn}(\varepsilon_{p,\rm max}),~t'_{ p\gamma}(\varepsilon_{p,\rm max})]=t'_{\rm acc}(\varepsilon_{p,\rm max})$,
we can estimate the maximum energy of the accelerated protons $\varepsilon_{p,\rm max}$.
As shown in Figure \ref{fig:timescales}, we have 
$t_{\rm syn}(\varepsilon_{p,\rm max})=t'_{\rm acc}(\varepsilon_{p,\rm max})<t'_{\rm dyn}<t'_{ p\gamma}(\varepsilon_{p,\rm max})$, 
and the maximum energy of accelerated protons is 
\begin{eqnarray}\label{Epmax}
\varepsilon_{p,\rm max}=1.2\times10^{21}{\rm eV}
(1+z)^{-1}\Gamma_{2.7}^{3/2}\eta_{0}^{1/2}\epsilon_{e,-1}^{1/4}   \epsilon_{B,-2}^{-1/4}R_{15}^{1/2}L_{\gamma,54}^{-1/4},
\end{eqnarray}
with $\eta$ as the acceleration efficiency.

{ Protons with lower energy will be confined in the acceleration site due to the strong magnetic field.
Those protons with the Larmor radius $R'_L(\varepsilon_p)$ and the cooling scale of the photomeson interaction $ct'_{p\gamma}(\varepsilon_p)$ larger than the shell thickness $\Delta r'$, would be able to escape directly.
The escape rate of protons with energy of $\varepsilon_p$ is estimated as \citep{Baerwald2013}
\begin{equation}\label{fesc}
f_{\rm esc}=\lambda'_{\rm mfp}(\varepsilon_p)/\Delta r'={\rm min}[\Delta r', R'_L(\varepsilon_p), ct'_{p\gamma}(\varepsilon_p)]/\Delta r'
\end{equation}
where $\lambda'_{\rm mfp}(\varepsilon_p)\equiv {\rm min}[\Delta r', R'_L(\varepsilon_p), ct'_{p\gamma}(\varepsilon_p)]$ is the mean free path of protons with energy of $\varepsilon_p$.
The shell thickness is $\Delta r'\simeq \Gamma c t_{\rm v}/(1+z)=1.2\times10^{12}{\rm cm}~(1+z)^{-1}\Gamma_{2.7}\frac{t_{\rm v}}{0.082{\rm s}}$, the Larmor radius of protons is $R'_L(\varepsilon_p)=\varepsilon_p(1+z)/(\Gamma\eta e B')=1.3\times10^{12}{\rm cm}~(1+z)\varepsilon_{p,21}\eta_0^{-1}\epsilon_{e,-1}^{\frac{1}{2}}
    \epsilon_{B,-2}^{-\frac{1}{2}}R_{15}L_{\gamma,54}^{-\frac{1}{2}}$, and $e$ as the elementary charge.
For protons with energy $\varepsilon_{\rm p}>\varepsilon_{\rm p,esc}= 9.6\times10^{20}{\rm eV}(1+z)^{-2}\Gamma_{2.7}\eta_0\epsilon_{e,-1}^{-\frac{1}{2}}
\epsilon_{B,-2}^{\frac{1}{2}}R_{15}^{-1}L_{\gamma,54}^{\frac{1}{2}}\frac{t_{\rm v}}{0.082{\rm s}}$, 
there are $\Delta r'< R'_{\rm L}$ and $\Delta r'<ct'_{p\gamma}$.
According to Eq. (\ref{fesc}), the escape efficiency is $f_{\rm esc}=1$ for protons with energy $\varepsilon_{\rm p}>\varepsilon_{\rm p,esc}$
\footnote{To be noted here, we only consider the energy loss of protons due to interactions with prompt photons but ignore possible interactions with afterglow photons. }.}

Such escaped protons 
will be deflected by magnetic fields in the host galaxy. 
The deflection caused by the regular magnetic field in the host galaxy is dependent on the GRB location in the host galaxy and the pointing direction of the GRB jet.
The Hubble space telescope observations reveal a disk-like host galaxy with
an effective radius of about $R_{\rm e}=2.45\pm0.20\,\rm kpc$ \citep{JWST2023}.
The host galaxy is viewed close to edge on with 
the burst about 0.65 kpc offset from the nucleus \citep{JWST2023}.
Therefore, 
deflections on protons by the regular magnetic field in the host galaxy is approximated to be \citep{Finley2006}
\begin{equation}\label{theta_reg_hg}
    \theta_{\rm reg, hg}\simeq4.1^\circ\varepsilon_{p,19.5}^{-1}\frac{B_{\rm reg,hg}}{{ 1\,\mu \rm G}}\frac{R_{\rm e}}{{\rm 2.45\,\rm kpc}},
\end{equation}
with $B_{\rm reg,hg}$ is the average strength of the regular magnetic field.
The root mean square (rms) of random deflections due to the turbulent magnetic field in the host galaxy is approximated as \citep{Lee1995,Harari2002,Finley2006}
\begin{equation}\label{theta_random_hg}
    \theta_{\rm rms,hg}\simeq 1.5^\circ \varepsilon^{-1}_{p,19.5}\frac{B_{\rm rms,hg}}{{2\mu \rm G}}\left(\frac{\lambda_{B,\rm hg}}{{\rm 50~pc}}\right)^{1/2}\left(\frac{L_{\rm hg}}{{10\rm~ kpc}}\right)^{1/2},
\end{equation}
where $B_{\rm rms,hg}$ and $\lambda_{B,{\rm hg}}$ are the rms strength and the coherent length of the turbulent magnetic field strength, and $L_{\rm hg}$ is the propagation length of protons in the turbulent magnetic field.
The rms of the time delay due to the turbulent magnetic field is approximated as \citep{Alcock1978,Batista2022}
 \begin{eqnarray}
   &&\Delta t_{\rm rms,hg}\simeq\frac{L_{\rm hg}\theta_{\rm rms, hg}^2}{12c}
   \simeq 1.9\, {\rm yr}\varepsilon^{-2}_{p,19.5}\left(\frac{B_{\rm rms,hg}}{2~\mu \rm G}\right)^2\frac{\lambda_{B,\rm hg}}{\rm 50\,\rm pc}\left(\frac{L_{\rm hg}}{{10\,\rm kpc}}\right)^2.   
 \end{eqnarray}

If only considering the deflection by the regular magnetic field and meanwhile $\theta_{\rm reg,hg}$ is larger than the jet half-open angle $\theta_{\rm j}$, protons escaping from the host galaxy will not point to the Milky Way (MW). 
Taking account of the deflection by the turbulent magnetic field in the host galaxy and in the extragalactic field, a small fraction of protons might still have a chance to arrive at the MW.
If 
$\theta_{\rm reg,hg}<\theta_{\rm j}$ and $\theta_{\rm rms,hg}\gg\theta_{\rm j}$,
the flux of protons pointing to Earth would be reduced significantly.
LHAASO's observation on the afterglow emission suggests a very narrow jet 
with a half-open angle $\theta_{\rm j}\simeq0.6^{\circ} E_{\rm k,55}^{-1/8}n_0^{1/8}$ \citep{LHAASO2023},
where $E_{\rm k}$ is the isotropic kinetic energy and $n$ is the circumstance density.
If assuming $E_{\rm k}=3\times10^{55}{\rm erg}$, $n=1$, $B_{\rm reg,hg}=1~\mu\rm G$, $B_{\rm  rms,hg}=2~\mu\rm G$, $L_{\rm hg}=10\rm~kpc$ and $\lambda_{B,{\rm hg}}=50\rm~pc$, for protons with energy $\varepsilon_p\gtrsim 300\,\rm EeV$, there are $\theta_{\rm reg, hg}<\theta_{\rm j}$ and $\theta_{\rm rms,hg}<\theta_{\rm j}$, indicating that protons with energy $\gtrsim300\,\rm EeV$ injected into the inter-galactic space from the host galaxy are still pointing to Earth.
The rms of the time delay for $\gtrsim 300\,\rm EeV$ protons is about 7 days, which will not affect the flux of detected protons in an observation time of a few years.
If magnetic fields of the host galaxy are weaker, the threshold energy of protons pointing to Earth might be lower than 300 EeV.
For the sake of conservation, we take account of protons with energy $\gtrsim 300$ EeV that directly escaping from the burst, as a component of UHECRs that injected into the inter-galactic space and propagating to the direction of Earth. 

\section{ The production and escape of neutrons} 
Free neutrons are produced via the photomeson production interaction in the burst, 
\begin{equation}
	p + \gamma \rightarrow \Delta^+ \rightarrow \left\{\begin{array}{lc} n + \pi^+    \\[0.2cm]  p + \pi^0.
	\end{array} \right.  
	\label{equ:Delta}
\end{equation}
and would decay into protons via the $\beta-$decay, i.e.,
\begin{equation}
n \rightarrow  p + e^- + \bar{\nu}_e,
\label{equ:ndec}  
\end{equation}
with a lifetime of about $t'_{n}\simeq 879.6~ {\rm s}$ in the rest frame \citep{workman2022}.

The timescale for the neutron production is
\begin{eqnarray}
t'_{p\gamma\rightarrow n}\simeq \frac{\kappa_{p\gamma}}{\rho\kappa_n\Re_n} t'_{p\gamma}\simeq4.3 t'_{p\gamma}.  
\end{eqnarray}
where $\rho=0.23$ is the cross-section ratio, $\kappa_n=0.4$ is the fraction of the proton energy converted into secondary neutrons, and $\Re_n=1/2$ is the neutron multiplicity, for the target photon energy in the proton rest frame being larger than $2000 \,\rm MeV$ \citep{Mucke2000,Mannheim2000,Dermer2012}.   
Therefore, a fraction of the proton energy is converted into the neutron energy.
The produced neutrons will lose energy via the $n\gamma$ interaction, with a cross section similar to that of $p\gamma$ interaction, i.e., $t'_{n\gamma}\simeq t'_{p\gamma}$.
The optical thickness for neutron escape is 
$\tau_{n}\simeq(t^{'-1}_{n\gamma}+t^{'-1}_{n})/t^{'-1}_{\rm dyn}
\simeq t'_{\rm dyn}/t'_{p\gamma}$\citep{Baerwald2013}.
In the case of GRB 221009A, 
neutrons can escape from the burst since $\tau_{n}< 1$. 

Ultra-high-energy neutrons with energy of $\varepsilon_n$ will be able to travel a distance of  
\begin{equation}\label{Ln}
    D_n=ct'_n\frac{\varepsilon_n(1+z)}{m_n}\simeq 88~{\rm kpc}~\varepsilon_{n,19}(1+z)
\end{equation}
before decaying into protons,  
which is considerably larger than the scale of most galaxies for $\varepsilon_n\gtrsim10^{19}{\rm eV}$,
where $m_n$ is the rest mass of neutrons. 
Therefore, neutrons with energy larger than 10 EeV, produced in the photomeson production interactions, are able to escape from the host galaxy with no deflections or time delay, and then decay into protons in the inter-galactic space. The neutron-decay-induced protons will be an important component of UHECRs that injected into the inter-galactic space.

\section{Spectra of accelerated protons, escaped protons and secondary particles}
Protons are assumed to be accelerated following a power-law spectrum with an exponential cutoff at the highest energy range,
\begin{equation}
\varepsilon_{\rm p}^2\frac{dN^{\rm acc}_{\rm p}}{d\varepsilon_{\rm p}} = A_p\varepsilon_{\rm p}^{-p+2}{\rm exp}(-\varepsilon_{p}/\varepsilon_{p,\rm max}),
\label{acceleratedp}
\end{equation}
where the normalization coefficient $A_p=\xi_p E_{\gamma,\rm iso}/{\rm ln}(\varepsilon_{p,\rm max}/\varepsilon_{p,\rm min})$ for $p=2$, 
$\varepsilon_{p,\rm min}$ is assumed to be the rest energy of protons. The luminosity of the total protons are $L_p=\xi_p L_\gamma=10^{54}{\rm erg}~\xi_{p,0}L_{\gamma,54}$.
The adopted values of $\xi_p$, $E_{\gamma,\rm iso}$, $L_{\gamma}$, $\varepsilon_{p,\rm max}$ and $\varepsilon_{p,\rm min}$ are listed in Table \ref{table_parameters}.

The spectrum of direct escaping protons can be approximated as
\begin{equation}
\varepsilon^2_{\rm p}\frac{dN^{\rm esc}_{\rm p}}{d\varepsilon_{\rm p}}=\varepsilon_p^2\frac{dN_p^{\rm acc}}{d\varepsilon_p}f_{\rm esc}=A_{p}{\rm exp}(-\varepsilon_{p}/\varepsilon_{p,\rm max})f_{\rm esc}.
\label{escapep}
\end{equation}

The production of secondaries from the photomeson production process are calculated via the following formula numerically,
\begin{equation}
    \frac{dN_s}{d\varepsilon_s}
    = \int \frac{d\varepsilon_p}{\varepsilon_p} \frac{dN_p^{\rm acc}}{d\varepsilon_p} \int d\varepsilon_\gamma \frac{dn_\gamma}{d\varepsilon_\gamma} \mathcal{R}(\varepsilon_s, \varepsilon_p),
    \label{neutrons}
\end{equation}
where $s$ denotes neutrons and neutrinos, and $\mathcal{R}(\varepsilon_s, \varepsilon_p)$ represents the production rate of secondary particles, including neutrons and neutrinos with energy $\varepsilon_s$, produced from primary protons with energy $\varepsilon_p$.
The production rate table $\mathcal{R}(\varepsilon_s, \varepsilon_p)$ can be generated with the Monte Carlo generator SOPHIA \citep{Sophia2000}.
Parameters adopted in the calculation are listed in Table \ref{table_parameters},
following measurements and constraints based on the multi-messenger observations.

In the calculation, the synchrotron cooling of the secondary pions and muons \citep{He2012} have been taken into account. 
Charged pions produced in the photomeson interaction will decay into 4 final state leptons, via
processes $\pi^{\pm} \rightarrow \nu_{\mu}(\bar{\nu}_\mu)\mu^{\pm}
 \rightarrow\nu_\mu(\bar{\nu}_\mu) e^+(e^-)\nu_e(\bar{\nu}_e)\bar{\nu}_\mu(\nu_\mu)$,
which approximately share the pion energy equally.
The all flavor neutrino spectrum can be calculated as \citep{He2012}
\begin{equation}\label{nunnutotaltheory}
\epsilon_{\nu} \frac{dn_{\nu}}{d\epsilon_{\nu}}\mathrm{d}\epsilon_{\nu}
= \frac{3\Re(\epsilon_p)}{4}f_{p\gamma}(\epsilon_p)\theta_\nu(\epsilon_p)
 \epsilon_p \frac{dn_p}{d\epsilon_p}\mathrm{d}\epsilon_p ,
\end{equation}
with the factor
$\theta_\nu(\epsilon_p)=\frac{1}{3}\zeta_\pi(\epsilon_p)+\frac{2}{3}\zeta_\pi(\epsilon_p)\zeta_\mu(\epsilon_p)$
accounting for the cooling of secondary
particles, 
where $\zeta_{\pi}=1-\exp{(-t_{\pi,\rm syn}/\tau_{\pi})}$
and $\zeta_{\mu}=1-\exp{(-t_{\mu,\rm syn}/\tau_{\mu})}$ are suppression factors caused by the synchrotron cooling of pions and muons,
and $\Re$
is the ratio between the amount of the charged pions and the total pions.
The synchrotron cooling timescales of pions and muons can be calculated as
$t_{\pi,\rm syn}=0.041~{\rm s}~(1+z)^{-1}\varepsilon_{\pi,18}^{-1}\epsilon_{e,-1}   \epsilon_{B,-2}^{-1}\Gamma_{2.7}^{3}R_{15}^{2}L_{\gamma,54}^{-1}$
and 
$t_{\mu,\rm syn}=0.015~{\rm s}~(1+z)^{-1}\varepsilon_{\mu,18}^{-1}\epsilon_{e,-1}    \epsilon_{B,-2}^{-1}\Gamma_{2.7}^{3}R_{15}^{2}L_{\gamma,54}^{-1}$,
and the life time of pions and muons can be written as
 $\tau_{\pi}=0.37{\rm~ s}\epsilon_{\pi,18}\Gamma_{2.7}^{-1}(1+z)$ and $\tau_{\mu}=41.5{\rm~ s}\epsilon_{\mu,18}\Gamma_{2.7}^{-1}(1+z)$.
 The energy of pions and muons in the observer's frame are approximated to be $\epsilon_{\pi}\simeq 0.2\epsilon_{\rm p}$ and $\epsilon_{\mu}=0.15\epsilon_{\rm p}$.

 The characteristic cutoff energy of pions and muons due to the synchrotron cooling, can be calculated via equaling the life time and the synchrotron cooling timescale, i.e.,
\begin{equation}
 \varepsilon_{\pi,\rm cut}=1.9\times10^{17}~{\rm eV}~(1+z)^{-1}\epsilon_{e,-1}^{1/2}
    \epsilon_{B,-2}^{-1/2}\Gamma_{2.7}^2R_{15}L_{\gamma,54}^{-1/2}
\end{equation}
and 
\begin{equation}
\varepsilon_{\mu,\rm cut}=1.0\times10^{16}~{\rm eV}~(1+z)^{-1}\epsilon_{e,-1}^{1/2}    \epsilon_{B,-2}^{-1/2}\Gamma_{2.7}^2R_{15}L_{\gamma,54}^{-1/2}.
\end{equation}

In Figure \ref{fig:neutrinos}, we plot the all-flavor neutrino fluence. 
The IceCube found none track-like events to the direction of the GRB from the fast-response analysis (FRA; \cite{IceCube2022GCN,IceCube2023}) in a time window of $-1$ hour to $+2$ hours from the {\it Fermi}-GBM trigger time, assuming a power law neutrino spectrum with different index p
\citep{IceCube2022GCN}. 
We plot the IceCube $90\%$ CL upper limit of all flavor neutrinos for index $p=1.5$ in Figure \ref{fig:neutrinos}, which shows that our predicted neutrino flux does not violate the IceCue upper limit,
assuming the flavor ratio as $1:1:1$ after oscillation.

{ The Giant Radio Array for Neutrino Detection (GRAND, \cite{GRAND2020}) is a scheduled detector for high energy particles with energy larger than $10^{17}\,{\rm eV}$, including neutrinos, UHECRs, and gamma-rays. 
The full planned configuration of GRAND will cover an area of $200,000\,{\rm km^2}$.
The GRAND sensitivity for transient neutrino flares with zenith angle of $90^\circ$ are plotted in Figure \ref{fig:neutrinos}.
Before taking account of the synchrotron cooling of pions and muons, the neutrino spectrum peaks around $10^{17}-10^{18}\,{\rm eV}$, which is just the sensitive band of GRAND. However, by taking account of the synchrotron cooling suppression, the neutrino fluence is suppressed in the sensitive band of GRAND.
Hence, if adopting the equilibrium energy of magnetic field $\epsilon_{B}=10^{-2}$, neutrinos from GRB 221009A-like event will not be able to be detected by GRAND, as shown in Fig. \ref{fig:neutrinos}. 
If a lower equilibrium energy of magnetic field is assumed, for instance, $\epsilon_{B}=10^{-4}$, the energy of neutrinos can reach the GRAND's threshold. 
Therefore, future observations of GRAND on neutrinos from other GRBs similar to GRB 221009A would constrain the magnetic field in the burst, and further constrain the acceleration mechanism of cosmic rays and the GRB central engine model.}

\section{Propagation in the inter-galactic space and in the Milky Way}
Protons injected into the inter-galactic space are contributed by two components, as mentioned above,
one component is protons directly escaping from the burst, with energy above 300 EeV, and the other is the neutron-decay-induced protons, with energy above 10 EeV.
Assuming that protons accelerated in the burst follows Eq. (\ref{acceleratedp}) with $p=2$, as shown by the dashed-dotted line in Figure \ref{fig:spectrum700Mpc},
the spectra of the directly escaping protons and the neutron-decay-induced protons
injected into the inter-galactic space would follow the dark orange dotted line and the purple dotted line in Figure \ref{fig:spectrum700Mpc}, respectively.

In the inter-galactic space, protons are deflected by the intergalactic magnetic field (IGMF) randomly. The corresponding deflection and time delay of protons depends on the strength of the IGMF. 
The rms of the deflection angle is estimated as \citep{Finley2006}
\begin{equation}
    \theta_{\rm rms,IG}=2.9\times10^{-6\circ}\varepsilon^{-1}_{p,19.5}B_{\rm IG,-16}\left(\frac{\lambda_{\rm IG}}{1\,\rm Mpc}\right)^{1/2}\left(\frac{D}{745\rm\, Mpc}\right)^{1/2},
\end{equation}
where $B_{\rm IG}$ and $\lambda_{\rm IG}$ are the rms strength and the coherence length of the IGMF, and $D$ is the distance between the source and the MW.
The time delay due to the IGMF is about \citep{Alcock1978,Batista2022}
 \begin{equation}
 \Delta t_{\rm IG} \simeq 16 \,{\rm s}\,\varepsilon^{-2}_{p,19.5}B^{2}_{\rm IG,-16}\frac{\lambda_{\rm IG}}{1\,\rm Mpc}\left(\frac{D}{745\rm\, Mpc}\right)^2.
\end{equation} 
If $B_{\rm IG}\lesssim 10^{-13}{\rm G}$ and $\lambda_{\rm IG}\lesssim 1{\rm Mpc}$, there are $\theta_{\rm rms,IG}\ll\theta_{\rm j}$ and $\Delta t_{\rm IG}<5{~\rm yr}$ for protons with energy larger than $10$ EeV, 
therefore, the fluence of protons arriving at the MW within 5 years would not be suppressed by the deflection of the IGMF significantly.
A $\sim 400$ GeV photon to the direction of GRB 221009A was detected by {\it Fermi}-LAT with a time delay of 0.4 days, suggesting an IGMF strength $B_{\rm IG}\sim4\times10^{-17}{\rm G}$ \citep{Xia2022},
which is consistent with constraints made from TeV blazars \citep{Dermer2011, 
Podlesnyi2022}. 
Adopting $B_{\rm IG}\sim4\times10^{-17}{\rm G}$, in the inter-galactic space,
protons with energy larger than 10 EeV are deflected by extremely small angles and short time delays,
which can be ignored for the detection of UHECRs on Earth.

{
A 3D simulation on protons propagating in the inter-galactic space is operated via the software {\sc CRPropa 3.2} \citep{Batista2016,CRPropa2022}, taking account of interactions with the EBL \citep{Gilmore2012} and CMB photons, including photo-pion productions, photo-disintegration, and electron-pair productions.
The direct escape protons and neutron-induced protons are injected from a source at a comoving distance $647$ Mpc to a spherical observer with a radius of $R_{\rm obs}=100~{\rm kpc}$
\footnote{The input comoving distance of GRB 221009A is calculated via taking a flat $\Lambda$CDM cosmology with $H_0=67.4~\rm km s^{-1} Mpc^{-1}$ and $\Omega_M=0.315$\citep{Planck2020}.}.
In the 3D simulation, we assume a Kolmogorov-type turbulent magnetic field with the rms strength of $B_{\rm rms}=4\times10^{-17}{\rm~G}$ and the coherence length of $l_c = 192\rm~kpc$.
We consider a grid of $256^3$ cells covering a total volume of $\sim (7.7\rm~Mpc)^3$. The grid is assumed to be periodically repeated to cover the whole simulation region.
The Cash-Karp method \citep{Cash1990} is adopted for simulating charged particle propagation in a magnetic field.
The simulation shows that protons with energy larger than $30$ EeV lose more than $50\%$ energy after propagating a comoving distance of 624.7 Mpc.
Therefore, the fluence of ultra-high-energy protons arriving at the edge of the MW is significantly suppressed. 
Most of the UHECRs from GRB 221009A lose their energy to $\lesssim 60$ EeV after propagating in the inter-galactic space.
The protons arriving at the edge of the MW peak around the energy of $30-40$ EeV with the spectra shown by solid lines in Figure \ref{fig:spectrum700Mpc}.
From Figure \ref{fig:spectrum700Mpc}, one can tell that, the contribution from neutron-decay-induced protons (the purple solid line) is dominated over the contribution from the direct escape protons (the dark orange solid line).}

{For the propagation of UHECRs in the MW, the energy loss is negligible, and the GMF model we adopt follows \cite{Jansson2012a,Jansson2012b}'s model, including the random large-scale and turbulent small-scale components.
We derive a map of protons arriving at Earth following the probability distribution of UHECRs after propagating in the Galactic magnetic field (GMF), 
via applying Galactic lenses in {\sc CRPropa 3.2} \citep{CRPropa2022} on protons that arrive at the edge of the MW.
In Figure \ref{fig:lensedmapGRB}, we plot the map of 1000 UHECRs observed on Earth,
with the energy spectral shape of UHECRs at the edge of the MW following 
the summed spectrum of the two solid lines shown in Figure \ref{fig:spectrum700Mpc}.
We backtrack anti-protons from Earth to the edge of the MW at 20 kpc adopting the Cash-Karp propagation method \citep{Cash1990} via {\rm{\sc CRPropa 3.2}}, to measure the time delay $\Delta t$ caused by the \cite{Jansson2012a,Jansson2012b} GMF model and the fraction of protons that arriving at Earth within a time delay $f(\Delta t)$.
The time delay $\Delta t$ and deflection angles $\theta$ of those protons are plotted in 
Figure \ref{fig:timedelay}, which indicates that,
as time increases, protons arriving at Earth have a tendency to distribute around the direction of the GRB within a larger deflection angle.
The solid angle $\Omega_{\rm max}(\Delta t)$ is calculated corresponding to the circle region around the GRB with a radius of the maximum deflection angle $\theta_{\rm max}(\Delta t)$.
Values of $f(\Delta t)$ and the expected solid angle $\Omega_{\rm max}(\Delta t)$ for the time delay $\Delta t$ derived via our simulations on backtracking $2\times10^6$ anti-protons in the GMF, are listed in Table \ref{table_theta}.
In reality, the solid angle could be measured from the detected excess of UHECRs around the direction of the GRB. 
}

\section{Results and Prospects}
The expected counts of protons with the energy of $>10$ EeV from the GRB detected by October of the year $2022+\Delta t$ can be calculated as 
\begin{equation}
N_{\rm s}(\Delta t)=163 f(\Delta t)\xi_{p,0}\frac{A_{\rm eff}}{310 \rm~km^2}.
\end{equation}
where $A_{\rm eff}$ is the annual exposure of the detector.

We assume the background spectrum is the same as the spectrum of the isotropic UHECRs detected by the PAO. 
The expected UHECR background counts within a solid angle $\Omega_{\rm max}(\Delta t)$ around the GRB integrated over time by October of the year $2022+\Delta t$ can be estimated via
\footnote{Here we calculate the integration adopting half-a-year bins.}
\begin{equation}
    N_{\rm b}(\Delta t)=A_{\rm eff}\int^{\Delta t}_0\int^{\infty}_{10\rm EeV}J_{\rm CR}(\varepsilon
    _{\rm CR})\Omega_{\rm max}(dt)d\varepsilon_{\rm CR} dt,
\end{equation}

where $J_{\rm CR}(\varepsilon_{\rm CR})$ follows the spectrum of isotropic UHECRs detected by the PAO, as described by Eq. (8) in \cite{Auger2020}, in units of $\rm km^{-2}sr^{-1}yr^{-1}EeV^{-1}$.

The statistical significance of the UHECR excess can be calculated via the Li $\&$ Ma method \citep{Li1983,TA2014}
\begin{equation}
S_{\rm LM}=\sqrt{2}\left[N_{\rm on}{\rm ln}\left(\frac{2N_{\rm on}}{N_{\rm on}+N_{\rm off}}\right)+N_{\rm off}{\rm ln}\left(\frac{2N_{\rm off}}{N_{\rm on}+N_{\rm off}}\right)\right]^{1/2}
\end{equation}
where $N_{\rm on}=N_{\rm s}+N_{\rm b}$ and $N_{\rm off}=N_{\rm b}$.

The Pierre Auger Observatory (PAO), Telescope Array (TA), and TA$\times$4 are the largest UHECR detectors in the world. The exposure of the cosmic ray detector depends on the declination of the arriving UHECRs. 
{According to the directional exposure for surface detectors of the PAO and TA shown in the left panel of Figure 1 in \cite{Biteau2019}, 
the averaged annual exposure of the PAO and TA for the GRB's declination $19.7^{\circ}$ is adopted as $\sim 310 {\rm~km^2}$ and $\sim 170 {\rm~km^2}$, with compiled dataset of events with energies $E>8.9~{\rm EeV}$ and $E>10~{\rm EeV}$, respectively. 
The exposure of fully completed TA$\times$4 is approximately 4 times the TA exposure, while so far about half of the full TA$\times$4 is completed \citep{Kido2023}.
Counts of GRB UHECRs that will be detected by the PAO and the full TA$\times$4, for the baryon loading factor $\xi_p=1$, are plotted in Figure \ref{fig:detectedcounts}. 
The significance of the PAO and TA$\times$4 detection peaks around 2030-2035, reaches $4.3~\sigma$ and $6.4~\sigma$, respectively, and then decreases since the solid angle of the arriving UHECRs increases, more background is included, as the time increases.
By October 2027, after 5-year observations, the PAO and the full TA$\times4$ can detect $\sim 17$ and $\sim 36$ GRB UHECRs (plus $\sim 5$ and $\sim 11$ background UHECRs) with energy larger than 10 EeV and the maximum deflection angle $\theta_{\rm max}$ of $\sim 5^\circ$ around the direction of the GRB.
The corresponding significance of the excess is estimated to be $\sim 3.4~\sigma$ and $5.0~\sigma$, respectively. If combining the PAO and the fully completed TA$\times$4 data, the combined significance would become higher, which would exceed 5 $\sigma$ by 2027. }

The annual exposure of GRAND for UHECRs from GRB 221009A is about 10 times the PAO according to Figure 10 in the GRAND white paper \citep{GRAND2020}. If the construction of GRAND can be completed before 2035, 
GRAND can detect $\sim148$ GRB UHECRs plus $\sim347$ background UHECRs with the energy of $>10\rm \, EeV$ around the GRB, with 8-year operation from the beginning of 2035 to the end of 2042, for $\xi_p=1$, with a significance of the excess as $\sim5.1~\sigma$. 
The projected significance of the UHECR excess measurable for these three detectors is shown in Figure \ref{fig:detectedcounts},
and the prospects are indeed promising.

In \cite{He2016}, a Monte Carlo Bayesian method is proposed to study the correlation between UHECRs and source candidates, via studying the spatial distribution of UHECRs with different energies. 
In the future, we can take into account not only the spatial distribution of UHECRs depending on the energy but also the temporal information of the arriving UHECRs, to further check the correlation between the detected UHECRs and the GRB at high statistics. 

{In this work, we adopt a high bulk Lorentz factor $\sim 500$, as suggested by LHAASO observations on the afterglow. With a typical variability timescale of $\sim 0.082$ s, as observed by {\it Fermi}-GBM, we then take an energy dissipation radius $R\sim 10^{15}{\rm cm}$, where protons are accelerated. A large energy dissipation radius and a high bulk Lorentz factor of the ejecta are helpful in enhancing the maximum energy of the accelerated protons, as indicated by Eq. (\ref{Epmax}), and further leads to a high flux of observed UHECRs.
Interestingly, a large energy dissipation radius of $\sim 10^{15}$ cm may challenge the standard fireball internal shock model unless just a small fraction of electrons have been accelerated otherwise the synchrotron radiation of the shocked electrons will not be able to peak at MeV energies. Instead, a large energy dissipation radius of $10^{15}-10^{16}$ cm is natural for a Poynting-flux-dominated GRB outflow \citep[e.g.,][]{2005ApJ...635L.129F,zhang2011,Kumar2015}.  
Indeed, recently, \cite{Zhang2023} suggested that the ejecta of GRB 221009A was composed of a narrow ($\sim 0.6^\circ$ half opening angle) Poynting-flux-dominated outflow composition and a broader matter-dominated jet wing.
In the future, a successful detection of the UHECR outburst from GRB 221009A will not only directly establish the GRB-UHECR connection, but also provide us a valuable chance to measure or stringently constrain the IGMF and the GMF, and key parameters of the GRB physics, such as the energy dissipation radius, the bulk Lorentz factor and the baryon loading factor of the jet, and further reveal the radiation mechanism and the central engine of the GRB.}

\begin{table}[!tbh]
\centering
\caption{\label{table_parameters} \textbf {Adopted Parameters of GRB 221009A.} 
}
 \medskip
\begin{tabular}{ccc}
      \hline
            \hline
	\textrm{Descriptions}&\textrm{Symbols}&\textrm{Values}\\ 
       \hline
   
	        Redshift 	&$z$ 	& 0.151 \\ 
			Dissipation radius 	&$R$  & $10^{15}{\rm cm}$  \\
			Low energy photon index
   \footnote{Parameters for the brightest emission from $T_0+225.024{\rm s}$ to $T_0+233.216{\rm s}$ in the second pulse are listed \citep{Frederiks2023}, with $T_0=47821.648{\rm s}$ UT (13:17:01.648) on 2022 October 9 as the Konus-WIND trigger time.} 	&$\alpha$  	& 0.76  \\
			High-energy photon index$^a$ &$\beta$  	& 2.13 \\
			 Minimum energy of photon spectrum$^a$ &$\varepsilon_{\gamma, \rm min}$  	& $20\,\rm keV$  \\
			 Maximum energy of photon spectrum$^a$ &$\varepsilon_{\gamma, \rm max}$  	& $10\,\rm MeV$  \\
			Peak energy of photon spectrum$^a$ 	&$\varepsilon_{\gamma, {\rm b}}$  	& $3\,\rm MeV$  \\
			Calibration luminosity\footnote{The luminosity at 20\,keV--10\,MeV.} 	&$L_\gamma$&$1\times 10^{54}\,\rm erg\,s^{-1}$  \\
			Bulk Lorentz factor 	&$\Gamma$  	& 500 \\			
			Baryon loading factor 	&$\xi_{\rm p}$  & 1\\
   			Proton index 	&$p$  	& 2   \\
       { Minimum energy of proton spectrum} 	&$\varepsilon_{\rm p, \rm min}$  	& $9.8\times10^{10}\,\rm eV$  \\
			{ Maximum energy of proton spectrum} 	&$\varepsilon_{\rm p, max}$  	& $1.2\times10^{21}\,\rm eV$  \\
			Fraction of magnetic field energy 	&$\epsilon_B$  & 0.01 \\
   Fraction of electron energy 	&$\epsilon_e$  & 0.1 \\
   Total radiation energy 	&$E_{\gamma,\rm iso}$  & $10^{55}\,\rm erg$ \\
     \hline
           \hline
\end{tabular}
\end{table}
\clearpage
\setcounter{figure}{0}

\begin{table}[!tbh]
\centering
\caption{\label{table_theta} \textbf {The time delay $\Delta t$, the fraction of arriving UHECRs $f(\Delta t)$, the solid angle $\Omega_{\rm max}(\Delta t)$, 
and the significance detected by the PAO and TA$\times$4 by the October of certain years, assuming the baryon loading factor $\xi_{\rm p}=1$.
}}
 \medskip
\begin{tabular}{ccccccc}
      \hline
            \hline
 	\textrm{Time}&$\Delta t$[yr]&$f(\Delta t)$&$\Omega_{\rm max}(\Delta t)$[sr]&PAO Significance[$\sigma$]&TA$\times$4 Significance[$\sigma$]\\ 
     \hline
    2025.10&3&3.5$\%$&0.013&1.9&2.8\\
    2027.10&5&10.2$\%$&0.020&3.4&5.0\\
    2029.10&7&15.5$\%$&0.027&4.0&5.8\\
    2031.10&9&19.4$\%$&0.034&4.2&6.2\\
    2033.10&11&22.7$\%$&0.040&4.3&6.3\\
    2035.10&13&25.7$\%$&0.046&4.3&6.4\\
    2037.10&15&29.8$\%$&0.059&4.1&6.1\\
     \hline
           \hline
\end{tabular}
\end{table}
\clearpage

\begin{figure}
\centering
\includegraphics[width=0.8\textwidth]{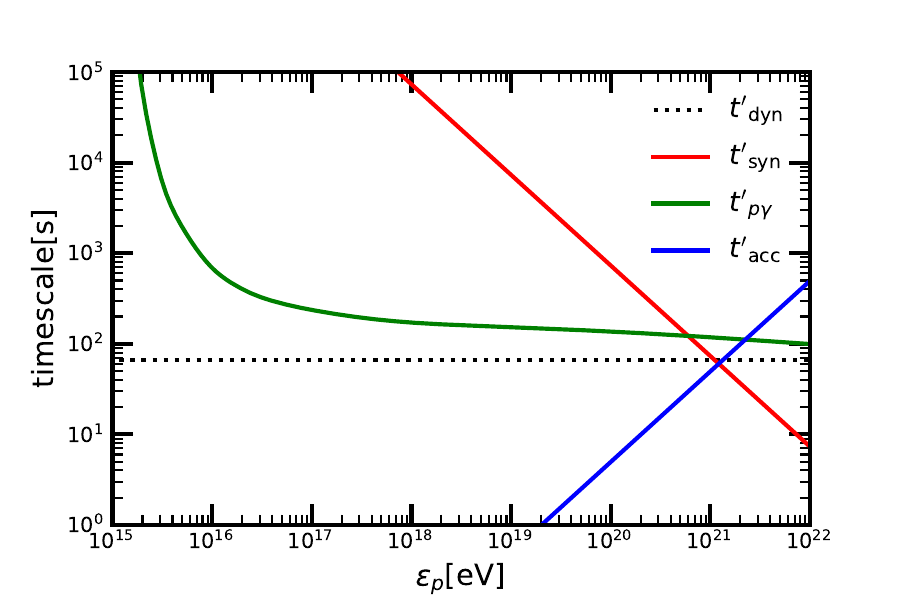}
\caption{\label{fig:timescales} { \bf Various timescales as functions of the proton energy.} All the timescales are measured in the comoving frame of the GRB outflow. The adopted parameters are listed in the Table \ref{table_parameters}.}
\end{figure}

\begin{figure}
\centering
\includegraphics[width=0.8\textwidth]{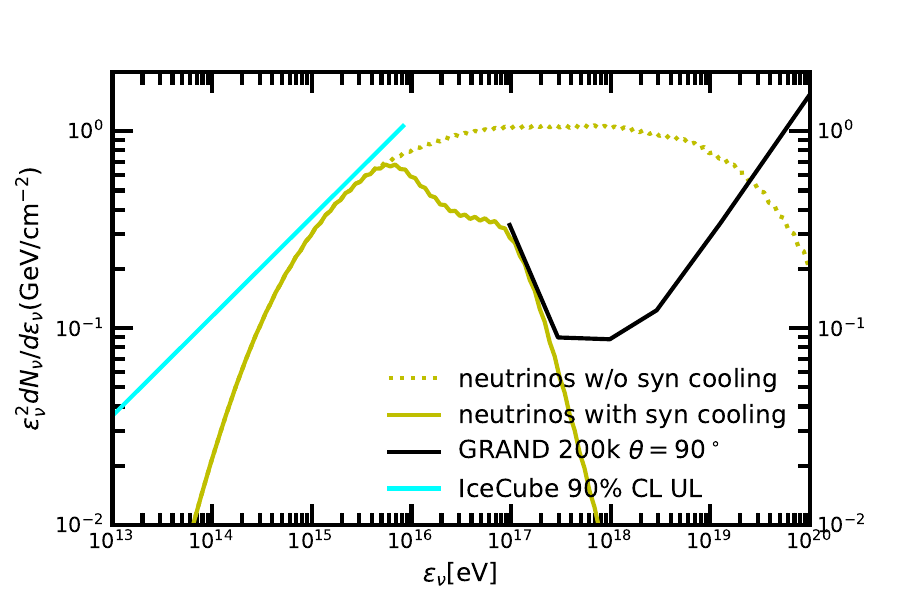}
\caption{\label{fig:neutrinos} {\bf The fluence of all flavor neutrinos from the source, adopting parameters as in Table \ref{table_parameters}.} The dark green dotted line is the neutrino spectrum without considering the synchrotron cooling of pions and muons, and the dark green solid line is the one considering the synchrotron cooling effect. The black solid line is the sensitivity of GRAND detecting neutrinos with a zenith angle of $\theta=90^\circ$ assuming null background.
The cyan solid line is the upper limit flux of all-flavor neutrinos in the IceCube's 3 hours FRA by assuming the spectral index as $p=1.5$.
}
\end{figure}

\begin{figure}[ht!]
\centering
\includegraphics[scale=0.8]{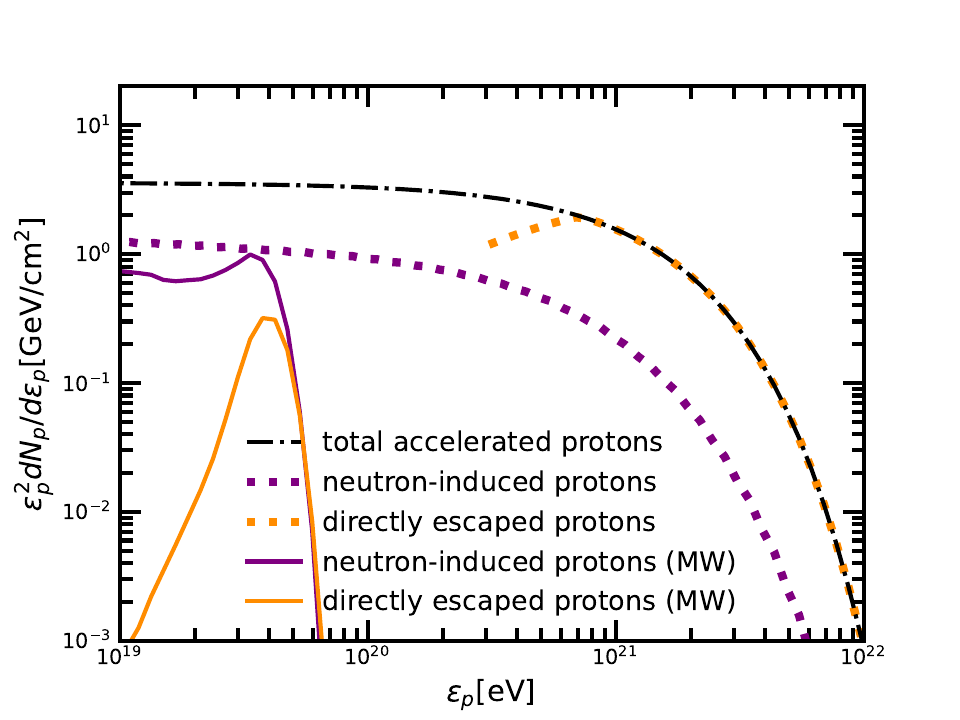}
\caption{{\bf The spectra of protons involved in this work.} The dash-dotted line, dotted lines, and solid lines represent protons accelerated in the burst, injected to the inter-galactic space, and arriving at the edge of the MW, respectively.
}
\label{fig:spectrum700Mpc} 
\end{figure}

\begin{figure}
\centering
\includegraphics[scale=0.8]{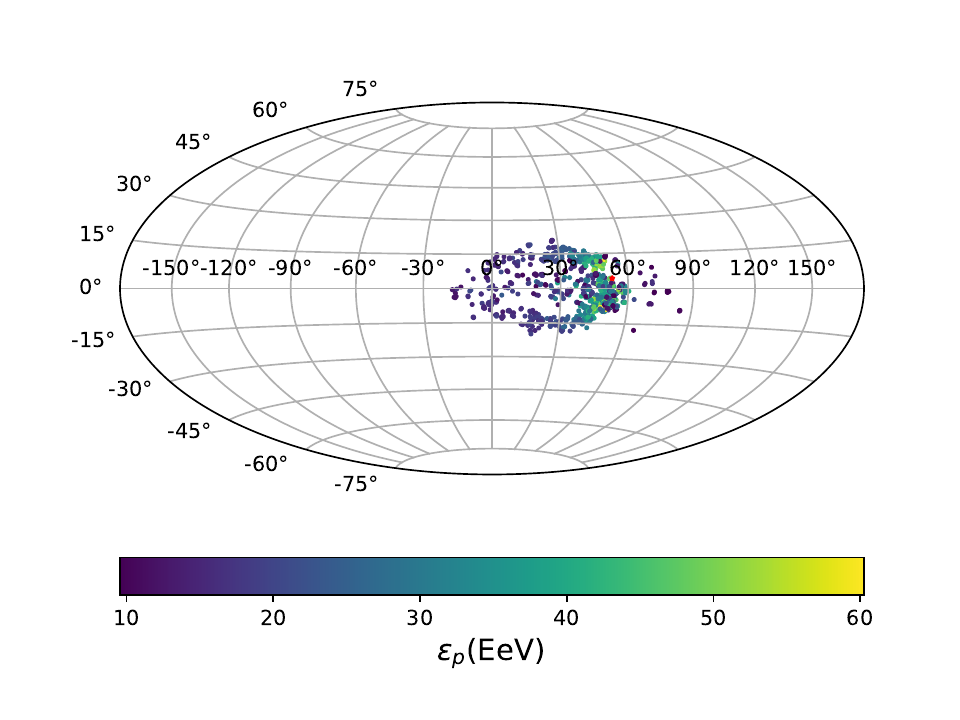}
\caption{{\bf The expected arrival distribution of 1000 UHECRs from GRB 221009A in the Galactic coordination, in the Hammer projection.} Colors represent energies of arriving UHECRs. The red star denotes the coordinate of GRB 221009A. 
}
\label{fig:lensedmapGRB} 
\end{figure}

\begin{figure}
\centering
\includegraphics[scale=0.8]{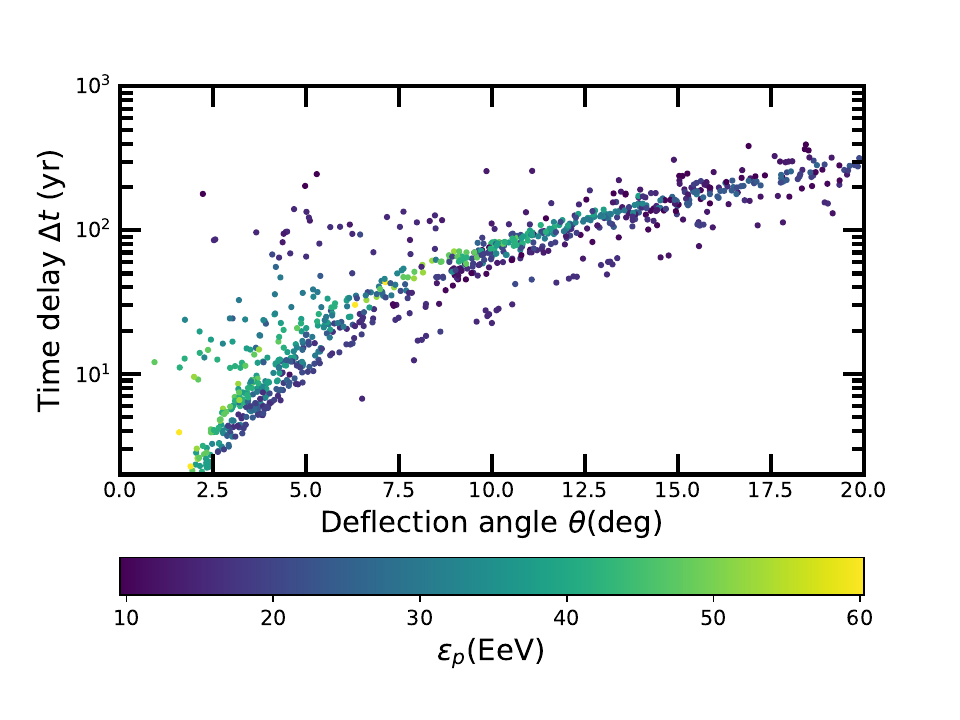}
\caption{{\bf The expected time delay and deflection angles of the arriving UHECRs from GRB 221009A.} 
The deflection and the time delay are mainly contributed by the Galactic magnetic fields. Colors represent energies of arriving UHECRs.    
}
\label{fig:timedelay} 
\end{figure}

\begin{figure}
\centering
\includegraphics[scale=0.9]{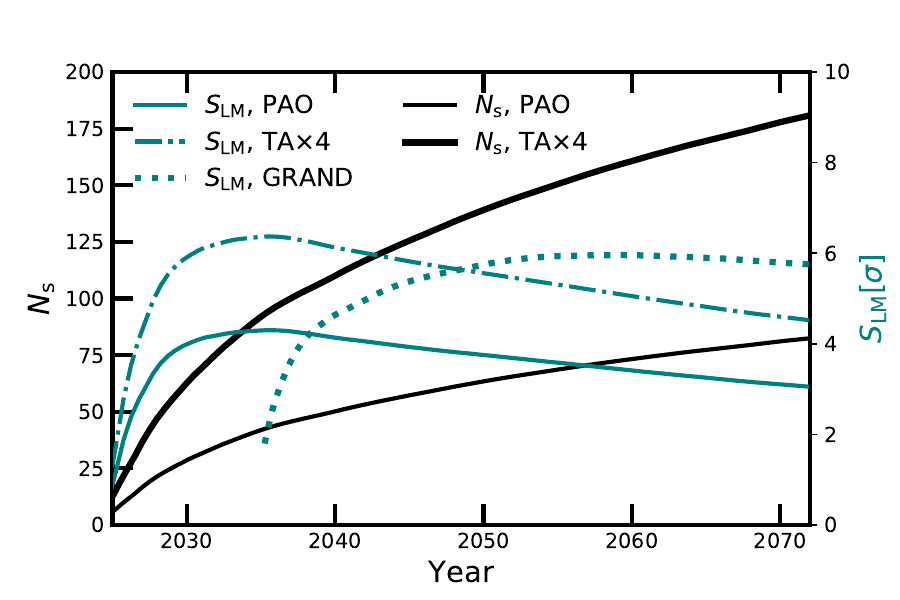}
\caption{{\bf The prospect of detecting the UHECR outburst from GRB 221009A.} The expected integral counts detected by the PAO and the full TA$\times$4, as a function of the time since the trigger time of GRB 221009A, are shown in solid lines. 
The teal lines denote the significance of the UHECR excess, detected by the PAO, the full TA$\times$4 and GRAND, respectively. The X-axis starts from the year of 2025.
 }
 \label{fig:detectedcounts}
\end{figure}

\begin{acknowledgments}
This work is supported by Project for Young Scientists in Basic Research of Chinese Academy of Sciences (No. YSBR-061), and by NSFC under the grants of No. 12173091, No. 11921003, No.12321003 and No. 12333006.
\end{acknowledgments}

\vspace{5mm}

\software{\sc{Astropy} \citep{2013A&A...558A..33A,2018AJ....156..123A},
          \sc{CRPropa 3.2} \citep{Batista2016},
          \sc{SOPHIA} \citep{Sophia2000}
}

\bibliographystyle{aasjournal}
\providecommand{\noopsort}[1]{}\providecommand{\singleletter}[1]{#1}%

\end{document}